# On the Generation of Multiply Charged Argon Ions in Nanosecond Laser Field Ionization of Argon Clusters

*Rajesh K. Vatsa,[a,*] and Deepak Mathur[b]*

[a] Chemistry Division, Bhabha Atomic Research Centre, Mumbai 400 085, India

[b] Department of Atomic and Molecular Physics, Manipal Academy of Higher Education, Manipal 576 104, India

[*] Corresponding Author; E-mail: rkvatsa@barc.gov.in

# ABSTRACT

Zhang and coworkers (*J. Phys. Chem. Lett.* **2020**, *11*, 1100−1105) have recently reported results of experiments involving irradiation of argon clusters doped with bromofluorene chromophores by nanosecond-long pulses of 532 nm laser light. Multiply-charged ions of atomic argon ($Ar^{n+}$, $1 \leq n \leq 7$) and carbon ($C^{n+}$, $1 \leq n \leq 4$) are observed which are sought to be rationalized using an evaporation model. The distinguishing facet of exploding clusters being progenitors of energetic ions and electrons constitutes the key driver for contemporary research in laser-cluster interactions; it is, therefore, important to point out inconsistencies that are intrinsic to the model of Zhang and coworkers. In light of similar reports already in the literature, we show that their model is of limited utility in describing the dynamics that govern how fast, multiply-charged atomic ions result from laser irradiation of gas-phase clusters. We posit that it is plasma behavior that underpins cluster heating and cluster explosion dynamics.

## TOC GRAPHICS

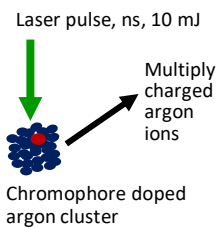

## KEYWORDS

Zhang and coworkers have recently reported results of experiments involving irradiation of aromatic chromophore (bromofluorene, $C_{13}H_9Br$) doped argon clusters by nanosecond-long pulses of 532 nm laser light.[1] The peak intensities of their laser pulses ($<10^{12}$ W cm$^{-2}$) were several orders of magnitude less than the intensities ($\sim 10^{16}$ W cm$^{-2}$) required to generate optical fields that match Coulombic fields within atoms. It is well established that the latter intensity conditions induce the formation of a nanoplasma within the irradiated cluster, with the multiply ionized constituents of the plasma undergoing ultrafast Coulomb explosion and, in consequence, yielding energetic and highly-charged atomic ions (for cogent reviews, see refs. 2-5, and references therein), Zhang et al. were, at intensity levels that were four orders of magnitude lower, able to observe multiply charged atomic ions of $Ar^{n+}$ ($1 \leq n \leq 7$) and $C^{n+}$ ($1 \leq n \leq 4$) upon irradiating argon clusters doped with molecules containing aromatic chromophores. Their attempt at rationalizing the observations were based on the fact that the doped aromatic molecule in their cluster has a resonance near 266 nm (achievable by two-photon excitation using 532 nm laser light). They postulated five-photon excitation/ionization of the dopant molecule (reached via a two-photon resonant state) followed by near-resonant energy transfer to the first excited Rydberg state of Ar which was, subsequently, ionized by absorption of another 532 nm photon. They proposed that the primary mode of energy absorption was resonant multiphoton absorption by the dopant molecule in their Ar-cluster. The solid-like cluster can accumulate a large number of charges in a very short period of time (<1 ps), thereby creating a strong Coulomb potential on the surface of the cluster, presumably in analogy with surface charges that have been observed in experiments with femtosecond laser pulses of $10^{16}$ W cm$^{-2}$ intensity.[6,7] Thermal evaporation of any surface atom or singly-charged argon ion in this Coulomb potential would result in the stripping of the electron back to the cluster, thus further ionizing the departing species and producing multiply charged atomic ions.

Does the rationalization offered by Zhang and coworkers[1] stand up to proper scrutiny vis-à-vis prevailing wisdom and, indeed, is their "model" of wider applicability? As is discussed in this Letter, we believe that the insights sought to be provided by Zhang and coworkers[1] are not consistent with what is already known about the dynamics of laser-cluster interactions, as exemplified in cogent reviews available in the literature.[2-5] It is important to point out inconsistencies that are intrinsic to the model of Zhang and coworkers, as we do in this Viewpoint, if for no other reason than the fact that the distinguishing facet of an exploding cluster being a progenitor of energetic ions and electrons constitutes a very important driver for contemporary research in the field.

The ramifications of the results of contemporary laser-cluster experiments are far-reaching. From a basic science viewpoint, clusters are, of course, intermediate to solids on the one hand and to atoms and small molecules on the other. The transition from single atom behavior to bulk properties is interesting and complex enough to merit investigation for its own sake. Upon irradiation of clusters that are not excessively large, a substantial percentage of constituent electrons undergo ionization; they depart the proximity of the cluster early in its break-up, and leave behind a hot, nonequilibrium, ionized plasma whose temporal evolution is strongly dependent on the optical radiation field. From a utilitarian perspective, clusters open up tantalizing new opportunities for developing table-top accelerators and sources of extreme ultraviolet radiation for lithography to be used in future generations of chip manufacturing technology. A great many other applications have been highlighted in the literature.[2-5]

Zhang and coworkers[1] correctly observe that energetic multiply-charged atomic ions have also been observed in a few other nanosecond, low-intensity experiments. Rationalization of the observations made in almost all such long-pulse experiments remains ambiguous, primarily because of the insistence on continuing invocation of the nanoplasma model that has been successfully used to describe the laser-cluster dynamics at intensity levels that are several orders of magnitude larger. At intensity levels of $10^{13}$ W cm$^{-2}$ and above, collisional inverse bremsstrahlung is the process that mainly underpins energy absorption by the irradiated cluster from the optical field that, consequently, results in multiple ionization and subsequent Coulomb explosion. The extrapolation of this model to the long-pulse regime is most certainly questionable because the ponderomotive energy is very much less than an eV at intensity levels used in long-pulse experiments; hence, it is doubtful whether ionized electrons can gain enough kinetic energy from the optical field alone to induce multiple ionization and acceleration of fragment atoms. Moreover, nanosecond timescales enable large-scale expansion of the irradiated clusters. Consequently, there is a rapidly decreasing electron-ion collision probability as the charged cluster expands to dimensions as large as a few microns on nanosecond time scales. In addition to these general observations, which the model of Zhang and coworkers does not address, we first focus on some particular issues that arise from their attempts to rationalize their observations.

Three crucial assumptions that lie at the core of the evaporation model of Zhang and coworkers[1] are (i) the requirement to have present within the cluster a chromophore (dopant aromatic molecule) that absorbs at 266 nm; (ii) the ability of the ionic state of this dopant to populate a specific Rydberg state of Ar; and (iii) generation of multiply charged argon ions by electron stripping as singly charged ions pass through the charged surface of the argon cluster.

Previous long-pulse experiments have shown the requirement (i) stipulated by Zhang and coworkers[1] is not at all necessary. For instance, in nanosecond experiments on multiple ionization of Ar-clusters doped with $Si(CH_3)_4$ (tetramethyl silane),[8] multiply charged argon ions up to +5 charge state were observed even at laser intensity levels below those used by Zhang and coworkers.[1] Tetramethyl silane does not possess a resonant state at 266 nm, making redundant the resonance condition of the dopant. Further, the cation $[Si(CH_3)_4]^+$ is not stable; it is known to dissociate on ultrafast timescales into $Si(CH_3)_3^+ + CH_3$,[9] thereby making it unlikely that it would contribute to any meaningful energy transfer into a Rydberg state of Ar. To illustrate that there is nothing special about the $Si(CH_3)_4$ dopant, consider experiments that have been reported in which iron pentacarbonyl $[Fe(CO)_5]$ is used as a dopant in Ar, Kr and Xe clusters. Upon irradiation with nanosecond pulses of 532 and 1064 nm light, highly charged ions of Ar, Kr and Xe have been observed with this dopant.[10] Once again, there is no resonance state in $Fe(CO)_5$. Thus, it appears improbable that resonances alone constitute the pathway to energy absorption by the cluster.

Assumption (ii) is that even the primary ionization of Ar-atoms is driven by energy transfer from the doped chromophore, with the ionic state of the dopant populating a specific Rydberg state of Ar. The sequence of events invoked by Zhang and coworkers is: excitation of the chromophore over and above the dissociation and ionization limits, followed by energy transfer to Ar-atoms, followed by creation of Rydberg-excited Ar atoms which are, subsequently, ionized by absorption of one more photon to finally create singly charged Ar-ions. The overall probability of these processes happening in sequence is, in our opinion, likely to be negligible for the following reasons. The chromophore used by Zhang and coworkers will most likely have an ionization energy of ~8-10 eV whereas the first Rydberg state of Ar lies at 11.7 eV. So, the excited neutral molecule will need to survive beyond both the dissociation limit (~ 2.9 eV for bromofluorene, $C_{13}H_9Br$, typical bond energy of C-Br) and the ionization threshold, until 11.7 eV of energy is accumulated for resonant collisional transfer to the lowest lying Rydberg state of the nearest Ar-atom. It is straightforward to utilize unimolecular dissociation RRK/RRKM theory[11] to show that the lifetime of an excited molecule at such high excess energies (11.7 – 2.9 eV= 8.8 eV or 850 kJ

mol$^{-1}$) over and above the dissociation threshold is negligible, making it unlikely in the extreme that any meaningful energy transfer can lead to the formation of Rydberg excited Ar-atoms.

Assumption (iii) is, in our opinion, most crucial; it offers the conjecture that generation of multiply charged argon ions is by electron stripping that occurs while Ar$^+$ traverse the charged surface of the cluster. Zhang et al.[1] have argued that about 1000 Ar$^+$ ions can be formed in their experiments in a matter of about 260 fs due to energy transfer. It is easily deduced that the low energy electrons, that are concomitantly generated during the ionization process, have a mean free path of ~4 nm; these electrons will, therefore, immediately leave the cluster (since the size of cluster is ~2 nm). Thus, the process of inner and outer ionization occurs simultaneously, leaving behind a charged cluster surface. A critical question that arises from such a scenario is: can a set of 1000 singly charged ions inside a cluster give rise to multiply-charged ions after some sort of charge redistribution that is induced by the Coulomb repulsion? To answer this, consider the work of Fennel and coworkers[12] who used an XUV beam (photon energy of 20-25 eV) of intensity 2 x 10$^{10}$ W cm$^{-2}$ to ionize Ar$_{3500}$. Due to the high photon energy and the relatively low intensity of their XUV pulse, Fennel et al.[12] showed that a reasonable fraction of argon atoms was ionized, giving rise to Ar$^+$, Ar$_2^+$, and some Ar$_3^+$. Furthermore, the high energy of ionizing photons would also ensure that the photoelectrons generated in their experiment would have a large kinetic energy and some, if not all, will leave the cluster, leaving its surface charged in similar fashion to that hypothesized by Zhang and coworkers.[1] Crucially, Fennel et al. were unable to observe any multiply charged argon ions in their experiments,[12] in contradiction to the expectations of the evaporation model of Zhang and coworkers. It is particularly noteworthy that in an extension of their initial experiment, Fennel et al. employed a two-pulse scheme wherein Ar-clusters were initially ionized by the XUV beam, and then subsequently irradiated by a near infrared beam which was temporally delayed by 5 ps. In such a scenario, multiply charged argon ions (up to 8+ charge state) were readily observed. It was found that the yield of the highest charged state ions depended sensitively upon the intensity of the IR laser pulse (3+ charge state was obtained at 2 x $10^{12}$ W cm$^{-2}$, rising to 6+ state at 4 x $10^{12}$ W cm$^{-2}$). These experiments unambiguously bring to the fore the crucial and sometimes subtle role of quasi-free electrons being efficiently heated by the NIR pulse, subsequently causing further ionization of Ar-ions by secondary electron impact. [Si(CH$_3$)$_4$] and Fe(Co)$_5$ each possess very different ionization states and a different set of lifetimes of individual cationic states. Despite this, the formation of multiply charged atomic ions is essentially the same for both molecules, as is the wavelength dependence (longer the wavelength, higher is the charge state).[8,10] What appears to be essential in the MCAI formation dynamics is the presence of quasi-free electrons in the nanoplasma and the optical field (either from longer laser pulses or from a second laser pulse, as in the case of Fennel's experiments).[12]

It is possible that in the experiments of Zhang et al.,[1] it is the rising part of their laser pulse that induces formation of ions and electrons. Thereafter, the ionized matter might possibly interact with the remaining portion of the long nanosecond laser pulse so as to cause secondary and tertiary ionization events. An easy way would have been for Zhang et al. to measure the electron energy in their experiments; such measurements would have helped clarify the overall dynamics.

We also note that, in the context of the energy transfer process suggested by Zhang and coworkers,[1] the rate of hole transfer in solid Ar is cited as 0.01 cm$^2$ V$^{-1}$ s$^{-1}$. The correct literature value is more than a factor of two larger (0.023 cm$^2$ V$^{-1}$ s$^{-1}$).[13] Moreover, the experiments from which this value is extracted were conducted on films which were 50-500 μm thick. By extrapolating over four orders of magnitude (to the dimensions of their cluster, 5-10 nm) Zhang and co-workers are asserting that such material properties are independent of system size!

So, is there an alternative theoretical underpinning to what is observed in long-pulse experiments on laser-cluster interactions? In contradiction to the expectations of the model put forward by Zhang and coworkers, we posit that it is plasma behavior that underpins cluster heating dynamics. It is now well established that electrons inside clusters can be direct absorbers of optical energy via collisional absorption. Strong enhancement in collisional absorption - of the order of as much as $10^3$-$10^4$ – is known to occur due to coherent superposition of collisions in clusters.[14] In the context of laser-cluster interactions, it is not only the laser intensity which determines the mechanism of ionization, laser frequency and pulse duration also play important roles and the latter parameter is not accounted for in the model presented by Zhang and coworkers.[1] In plasma physics, the quantity $I\lambda^2$ (I denotes the laser intensity and $\lambda$ is the wavelength of laser light) is defined as laser irradiance and many effects in laser-plasma interactions depend upon the magnitude of this quantity. An immediate implication is that threshold intensity for a given phenomenon can vary depending on the laser wavelength.[14] Depending on the density of electrons that are generated by photoionization inside the irradiated cluster, the nanoplasma that is generated upon laser irradiation can be underdense or overdense, and this is reflected in its frequency dependence.[3] The energy that is absorbed by an underdense plasma exhibits inverse square frequency dependence (or, equivalently, square wavelength dependence) whereas the energy absorbed in an overdense plasma is independent of frequency. Earlier work with nanosecond lasers has shown that the amount of energy being absorbed by plasma electrons increases with laser wavelength, eventually resulting in formation of higher charged states of atomic ions being formed.[8,10,15,16]

We also wish to draw attention to an unexpected facet reported by Zhang and coworkers in their Supplementary Material.[1] It appears that the externally applied DC field (used to extract ions from the laser-cluster interaction zone) is immaterial as the same charge distributions are obtained in presence of an external DC field as well as while using delayed extraction. So, the field that appears to be of primary importance in the experiments of Zhang and coworkers is that due to the presence of ions. If this is, indeed, the case, there is a need to deduce information about the number of ionic charges that are created, and their relative spatial distribution within the cluster, in order to calculate a value for the effective field. In the absence of this information, we have estimated the intra-cluster ionic charge in the following terms. In the case of a Xe-cluster of about 10 nm diameter, the cluster volume is $5 \times 10^{-19}$ cm$^3$. Within the nanoplasma model, the electron density at a laser frequency of $5.6 \times 10^{14}$ Hz (corresponding to 532 nm light) is $A/(5 \times 10^{-19})$. This yields a total of 1000-2000 electrons to enable the plasma frequency to be in resonance with the laser frequency. Taking the average number of atoms in the Xe cluster to be ~11000, it becomes clear that only 10-20% ionization can be expected assuming single ionization per Xe atom. This reduces to only 2-5% ionization if the average charge state of the Xe ion is taken to be 4-5 (as is indicated in the ion charge state distributions presented in reference 16 and also reported by Zhang et al.). It remains unclear how an evaporating singly charged ion – as postulated in their model – can give rise to the abundance of multiply charged ions observed in their experiments.

By way of a general consideration, we note that it would, of course, be incorrect to assert that intermediate states never play any role in the overall dynamics in laser interactions with doped clusters. For instance, in the case of molecular clusters of $CH_3I$ it has been shown that direct excitation of pure $CH_3I$ clusters occurs via a long-lived Rydberg *C* state that acts as an intermediate reservoir state for the ionization step.[11] This state, which has been well documented by UV and VUV spectroscopy, has long lifetime and high cross section (the latter being typical of Rydberg states) which readily enables excitation of the methyl iodide cluster upon 532 nm excitation. Nanosecond experiments have yielded a power dependence of ~3; formation of $CH_3I^+$ was also seen when only monomers were irradiated by 532 nm laser beam[17]. Thus, 3-photon excitation to a long-lived Rydberg state followed by another 2-photon absorption by the excited molecule leading to primary ionization has been clearly demonstrated in the case of $CH_3I$ clusters without the need to invoke energy transfer between molecules or cluster constituents or even electronic collisions.

**AUTHOR INFORMATION**

Deepak Mathur, Department of Atomic and Molecular Physics, Manipal Academy of Higher Education, Manipal 576 104, India; orcid.org/0000-0002-2322-6331;

Email: atmol1@gmail.com

**Corresponding Author**

Rajesh K. Vatsa, Chemistry Division, Bhabha Atomic Research Centre, Mumbai 400 072, India; orchid.org/ 0000-0002-9928-1572

Email: rkvatsa@barc.gov.in

**Notes**

The authors declare no competing financial interest.

## ACKNOWLEDGMENTS

D.M. thanks the Science and Engineering Research Board for financial support through the J. C. Bose National Fellowship (SR/S2/JCB-29/2006).## REFERENCES